# Atomic-scale investigation of γ-Ga$_2$O$_3$ deposited on MgAl$_2$O$_4$ and its relationship with β-Ga$_2$O$_3$

Jingyu Tang,[1,a)] Kunyao Jiang,[1] Chengchao Xu,[2] Matthew J. Cabral,[1] Kelly Xiao,[1] Lisa M. Porter[1] and Robert F. Davis[1,2]

**Affiliations**

[1]Department of Materials Science and Engineering, Carnegie Mellon University, Pittsburgh, Pennsylvania 15213, USA

[2]Department of Electrical and Computer Engineering, Carnegie Mellon University, Pittsburgh, Pennsylvania 15213, USA

[a)]Author to whom correspondence should be addressed: jingyuta@andrew.cmu.edu

**Abstract**

Nominally phase-pure γ-Ga$_2$O$_3$ was deposited on (100) MgAl$_2$O$_4$ within a narrow temperature window centered at ~470 °C using metal-organic chemical vapor deposition (MOCVD). The film deposited at 440 °C exhibited either poor crystallization or an amorphous structure; the film grown at 500 °C contained both β-Ga$_2$O$_3$ and γ-Ga$_2$O$_3$. A nominally phase-pure β-Ga$_2$O$_3$ film was obtained at 530 °C. Atomic-resolution scanning transmission electron microscopy (STEM) investigations of the γ-Ga$_2$O$_3$ film grown at 470 °C revealed a high density of antiphase boundaries. A planar defect model developed for γ-Al$_2$O$_3$ was extended to explain the stacking sequences of the Ga sublattice observed in the STEM images of γ-Ga$_2$O$_3$. The presence of the 180° rotational domains and 90° rotational domains of β-Ga$_2$O$_3$ inclusions within the γ-Ga$_2$O$_3$ matrix is discussed within the context of a comprehensive investigation of the epitaxial relationship between those two phases in the as-grown film at 470 °C and the same film annealed at 600 °C. The results led to the hypotheses that (i) incorporation of certain dopants including Si, Ge, Sn, Mg, Al, and Sc, into β-Ga$_2$O$_3$, locally stabilizes the "γ-phase" and (ii) the site preference(s) for these dopants promotes the formation of the "γ-phase" and/or γ-Ga$_2$O$_3$ solid solutions. However, in the absence of such dopants, pure γ-Ga$_2$O$_3$ remains the least stable Ga$_2$O$_3$ polymorph, as indicated by its very narrow growth window, lower growth temperatures relative to other Ga$_2$O$_3$ polymorphs, and the largest calculated difference in Helmholtz free energy per formula unit between γ-Ga$_2$O$_3$ and β-Ga$_2$O$_3$ than all other polymorphs.

**INTRODUCTION**

Beta-gallium oxide (β-Ga$_2$O$_3$) is an emerging ultra-wide-bandgap semiconductor material for applications in high-power electronic devices. Its higher theoretical breakdown electric field strength (~8 MV/cm)[1] surpasses that of SiC and GaN due to its larger bandgap of ~4.8 eV.[2–4] It also has significant advantages in bulk material production that uses cost-efficient melt growth techniques and in the growth of homoepitaxial thin films with controllable n-type doping using Si, Ge and Sn.[5–10] The successful incorporation of deep acceptors such as Fe and Mg[11–13] into β-Ga$_2$O$_3$ has facilitated the creation of semi-insulating regions that effectively reduce the leakage current of the device. Aluminum alloyed β-Ga$_2$O$_3$ has also been explored for two purposes (i) band gap engineering to further extend the breakdown electric field strength for high-power electronic devices[14,15] and (ii) the formation of a two-dimensional electron gas at the $(Al_xGa_{1-x})_2O_3/Ga_2O_3$ interface[16–19] for high electron mobility transistors. In addition to the applications in power diodes and transistors,[1,20] the bandgap energy of β-$Ga_2O_3$ falls in the deep ultraviolet range, so it is also a candidate material for solar-blind UV photodetectors.[21–23]

$Ga_2O_3$ exhibits five polymorphs which are α (trigonal), β (monoclinic), γ (cubic-defective spinel), δ (cubic-bixbyite) and κ(ε) (orthorhombic). It was earlier suggested that the δ-phase was a nanocrystalline form of the κ(ε)-phase.[24] However, recently reported research[25] has demonstrated the epitaxial growth of δ-$Ga_2O_3$ on a cubic bixbyite-structured $Fe_2O_3$ buffer layer deposited on an yttrium-stabilized zirconia substrate. The exact crystal structure of δ-$Ga_2O_3$ remains a subject of ongoing debate; the structures of the other four phases are universally accepted. The β-phase is the thermodynamically stable polymorph from room temperature to its melting point. However, two or more polymorphs have been determined to occur in selective $Ga_2O_3$ thin films.[26–30] For example, γ-phase inclusions have been observed in Si, Ge, and Sn doped and Al and Sc alloyed β-$Ga_2O_3$ films as well as bulk β-$Ga_2O_3$ crystals.[31–38] A highly defective $Ga_2O_3$ intermediate layer with a close structural arrangement to γ-$Ga_2O_3$ was determined to have formed between a Au contact and a (100) β-$Ga_2O_3$ film,[39] affecting the electrical characteristics of the film. An understanding of the impact of "γ-phase" inclusions on the electronic and thermal properties of β-$Ga_2O_3$ and their elimination are crucial for the repeated fabrication of devices with replicable properties.

γ-$Ga_2O_3$, an analog of γ-$Al_2O_3$ (space group: $Fd\bar{3}m$), has been surmised to have a cation defective spinel structure.[40,41] As such, 8/3 vacancies must be introduced into the cation sites of each unit cell to maintain the stoichiometric ratio between Ga and O at 2:3. However, the uncertainty of the preferred sites for the vacancies[42–45] in addition to the deviation from the ideal spinel structure associated with partial filling of non-spinel positions (e.g., the 16c and 48f Wykoff positions[24,43]) by Ga atoms make this polymorph even more complicated. In terms of first-principles calculations, the Helmholtz free energies per formula unit for $Ga_2O_3$ polymorphs are in the sequence of β < κ(ε) < α < γ,[45] indicating that γ-$Ga_2O_3$ is the least stable polymorph. Most of the γ-$Ga_2O_3$ materials synthesized by solution-based methods[40,46–48] suffer from poor crystallinity, small particle size, and the unavoidable contamination of β-$Ga_2O_3$, which adds to the difficulty of characterizing this polymorph. Some progress on epitaxial stabilization of γ-$Ga_2O_3$ thin films grown via mist-CVD, pulse laser deposition (PLD) and MOCVD[44,49–53] on substrates such as $Al_2O_3$, MgO and $MgAl_2O_4$ have been reported, but detailed atomic-scale investigations of defects within the film and their relationships to β-$Ga_2O_3$ have been rarely investigated.[32,36,44]

In this paper, the parameters employed and the results for the repeated growth of nominally phase pure γ-$Ga_2O_3$ films on (100) $MgAl_2O_4$ via MOCVD at 470 °C are presented. The temperature range for the deposition of γ-$Ga_2O_3$ was between 440 °C and 500 °C. Films grown at 530 °C were nominally phase pure β-$Ga_2O_3$. Previously, we reported the results and conclusions of our investigations of film growth and annealing within the higher temperature range of 550 °C to 1000 °C. In these studies, significant Mg and Al diffused from the $MgAl_2O_4$ substrates into the β-$Ga_2O_3$ layer. This resulted in the formation of γ-$Ga_2O_3$ solid solutions with a spinel structure which were referred to as the "γ-phase".[54] To avoid ambiguity, in this paper, we specifically use the term "γ-$Ga_2O_3$" to distinguish it from the "γ-phase", as no measurable diffusion of Mg and Al was determined via energy dispersive x-ray spectroscopy (EDX) to occur during the film growth. In-depth atomic-scale investigations of select γ-$Ga_2O_3$ films using STEM revealed the presence of significant antiphase boundaries. An analysis of the epitaxial relationship in this study between γ-$Ga_2O_3$ and β-$Ga_2O_3$ is discussed.

## METHODS

Single-side, chemo-mechanically polished, epi-ready (100) MgAl$_2$O$_4$ substrates were used for film growth in a vertical, low-pressure, cold-wall MOCVD reactor. The parameters employed for these films are detailed in Table I. Triethylgallium (TEGa) stored in a stainless-steel bubbler from Sigma-Aldrich and ultra-high purity O$_2$ (5N) were used as the reactants. The carrier gas for TEGa was ultra-high purity N$_2$ (5N), which also served as a diluent gas to prevent any pre-reaction of the precursors. For all the growths, the bath temperature and the bubbler pressure were kept at 20 °C and 150 Torr, respectively; thus, the TEGa flow rate was solely determined by the carrier gas flow rate. To maintain a constant O/Ga (VI/III) ratio of 635, the molar flow rates of O$_2$ and TEGa were kept at 0.0223 mol/min and 7.02×10$^{-5}$ mol/min, respectively. The reactor pressure was controlled using a butterfly valve and held at 20 Torr for all growths. The rotational speed of the Hastelloy susceptor was selected to be 120 revolutions per minute (rpm) to achieve a uniform flow over the substrates. The period for all growths was set to one hour. High-resolution x-ray diffraction (HRXRD) patterns were acquired using a Panalytical X'pert Pro MPD x-ray diffractometer (4-crystal Ge (220) monochromator) equipped with a Cu K$_{\alpha 1}$ (1.5406 Å) x-ray source. The film thickness measurements were conducted using a Filmetrics F50-200 spectrometer. Transmission electron microscopy (TEM) and STEM investigations including SAED, high-angle annular dark field (HAADF)-STEM imaging and STEM-EDX spectroscopy were performed using a Thermo Fisher Themis 200 G3 aberration-corrected TEM/STEM equipped with a Super-X EDX system. TEM lamella preparation for the γ-Ga$_2$O$_3$ films was conducted using an FEI Helios Xenon Plasma Dual Beam Focused Ion Beam (FIB).

TABLE I. Parameters for growth of Ga$_2$O$_3$ films on (100) MgAl$_2$O$_4$ and discussed herein.

| Exp.# | Temperature (°C) | Chamber pressure (Torr) | O$_2$ flow rate (sccm) | Carrier gas flow rate (sccm) | TEGa flow rate (sccm) | VI/III ratio | Growth rate (nm/h) |
|---|---|---|---|---|---|---|---|
| Exp.152 | 440 | 20 | 500 | 45 | 1.57 | 635 | N/A |
| Exp.130 | 470 | 20 | 500 | 45 | 1.57 | 635 | 407 |
| Exp.133 | 500 | 20 | 500 | 45 | 1.57 | 635 | 607 |
| Exp.131 | 530 | 20 | 500 | 45 | 1.57 | 635 | 649 |

## RESULTS AND DISCUSSION

### I. Investigation on the growth of γ-Ga$_2$O$_3$ on (100) MgAl$_2$O$_4$

In this section, a comprehensive suite of XRD scanning techniques including out-of-plane 2θ-ω scans, ϕ-scans, ω-scans, pole figure and reciprocal spacing mapping (RSM) were used to investigate (i) the growth window for γ-Ga$_2$O$_3$ grown on MgAl$_2$O$_4$, (ii) the film/substrate epitaxial relationship, and (iii) the lattice distortion/relaxation of the film. TEM-SAED patterns were acquired to further confirm the phase and crystal orientation of the deposited film. Figure 1(a) shows the out-of-plane 2θ-ω scans of the films grown using the parameters shown in Table I. The strongest peak at 44.79° indicated by the black star corresponds to the (400) planes of the MgAl$_2$O$_4$ substrate. The other peaks at 30.05°, 43.17°, and 44.14° are attributed to the (400) and ($\bar{1}$12) planes of β-Ga$_2$O$_3$[54] and the (400) plane of γ-Ga$_2$O$_3$[51], respectively. For the film grown at 440 °C, only the substrate peak was detected indicating that it was either poorly crystallized or amorphous. For the film grown at 470 °C, the (400) peak of γ-Ga$_2$O$_3$ was observed with no other film peaks, indicating an epitaxial relationship between the film and the substrate. The lattice mismatch between (400) γ-$Ga_2O_3$ and (400) $MgAl_2O_4$ is ~+1.9%; thus, these major peaks occur near each other. Further increases in the growth temperature to 500 °C and 530 °C resulted in the formation of (i) a mixture of β-$Ga_2O_3$ and γ-$Ga_2O_3$ and (ii) nominally phase pure β-$Ga_2O_3$, respectively. The analytical results for films grown at higher temperatures (550 °C-850 °C) under similar growth conditions have been presented and discussed in a separate paper.[54] As revealed in Fig. 1(a), the growth window of nominally phase pure γ-$Ga_2O_3$ was narrow and centered around 470 °C. Oshima et al. also reported a narrow growth window from 390 °C to 400 °C for pure γ-$Ga_2O_3$ deposited on (100) $MgAl_2O_4$ via mist-CVD.[51]

Figure 1(b) shows the φ-scan of the {440} γ-$Ga_2O_3$ grown at 470 °C acquired simultaneously with that of the $MgAl_2O_4$ substrate. The resulting four peaks evenly spaced at 90° intervals illustrate the four-fold symmetry of γ-$Ga_2O_3$ when rotated along the [100] direction. The absence of additional peaks indicates the absence of in-plane rotational domains of γ-$Ga_2O_3$. A comparison of the peak locations between the film and substrate reveals that the in-plane epitaxial relationship is [011] γ-$Ga_2O_3$ || [011] $MgAl_2O_4$, corresponding to cube-on-cube growth. The absence of out-of-plane rotational domains was verified through pole figure measurement of {444} γ-$Ga_2O_3$, as only four spots were detected at χ = 54.7°, as shown in Fig. 1(c). Both out-of-plane and in-plane epitaxial relationships between the film and substrate match the published results.[51,52,55] Distortion within the as-grown γ-$Ga_2O_3$ at 470 °C was quantified by measuring the full-width-at-half-maximum (FWHM) of the x-ray rocking curve (XRC) or ω-scan of both the (800) symmetric peak and the (440) asymmetric peak of γ-$Ga_2O_3$ shown in Fig. 1(d). The FWHM values of these two peaks were 0.718° and 0.720°, respectively, indicating that the mosaicity within the γ-$Ga_2O_3$ lattice is equally distributed both out-of-plane and in-plane. Figure 1(e) shows the RSM around the (804) diffraction spots of the film and substrate. The observation of horizontal separation between the film spot and the substrate spot in reciprocal space confirms that the volume of the film wherein diffraction occurred is in a relaxed state.

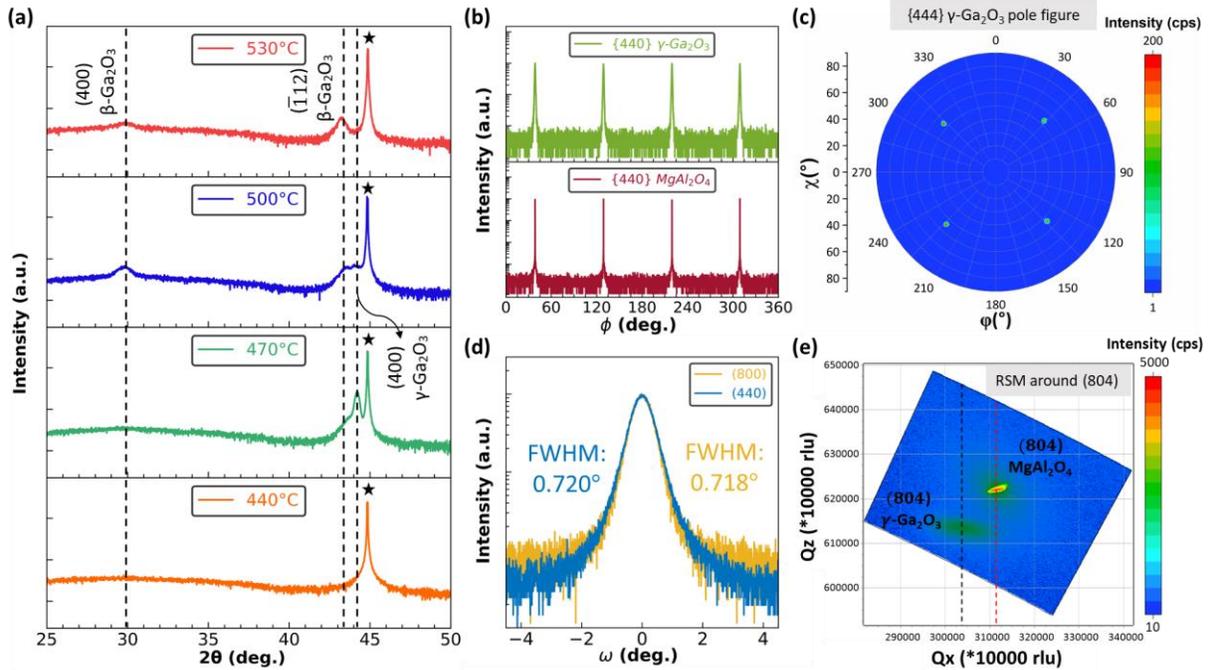

FIG. 1. (a) XRD patterns (log scale) for $Ga_2O_3$ films grown on (100) $MgAl_2O_4$ substrates. Black stars indicate the (400) reflection from the substrate. (b) XRD phi-scans of {440} γ-$Ga_2O_3$ (grown at 470 °C) and $MgAl_2O_4$, respectively. (c) Pole figure of diffraction from {444} γ-$Ga_2O_3$ grown at 470 °C. (d) XRD rocking curves of (800) symmetric and (440) asymmetric diffraction peaks of γ-$Ga_2O_3$ grown at 470 °C. (e) RSM around (804) diffraction spot of film/substrate combination (film grown at 470 °C).

For TEM/STEM analysis, a thicker film of ~700 nm was grown at 470 °C in two hours using the same parameters applied for the one-hour depositions. The peak positions in the XRD 2θ-ω pattern (not shown) for this film were the same as the one shown in Fig. 1(a); the intensity of the γ-$Ga_2O_3$ peak for the two-hour film increased due to its greater thickness. Figure 2 shows cross-sectional STEM and SAED patterns for the γ-$Ga_2O_3$ film grown at 470 °C acquired along the [010] and [011] zone axes, respectively. The film/substrate contrast was sharp, as can be observed in Figs. 2(a) and 2(d), and clearly differentiates the film from the substrate. SAED patterns of the film along both [010] (Fig. 2(b)) and [011] (Fig. 2(e)) zone axes matched well with the simulated SAED patterns of γ-$Ga_2O_3$ shown in Figs. 2(c) and 2(f). The crystal model for γ-$Ga_2O_3$ was built assuming $Fd\bar{3}m$ symmetry with fully occupied and well-ordered cubic close packing in the O sublattice lattice. The Ga sublattice was partially occupied on 8a and 16d Wyckoff sites to balance the stoichiometry.[40] Additionally, some elongated diffraction spots such as ($0\bar{2}2$), ($\bar{4}\bar{2}2$) and ($11\bar{1}$) indexed in red can be seen in Fig. 2(e). Other spots such as (400), ($\bar{2}\bar{2}2$) and ($04\bar{4}$) indexed in blue on the same figure possess a rounded shape without observable elongation. From structure factor calculations, the rounded diffraction spots are dominated by the well-ordered O sublattice while the elongated spots are dominated by the Ga sublattice. The degree of elongation is affected by the formation of planar defects within the Ga sublattice. Similar spot broadening has been reported in γ-$Al_2O_3$,[56] which is thought to have the same crystal structure as γ-$Ga_2O_3$. As will be seen in the next section, antiphase boundaries, a specific type of planar defect, were widely observed inside the γ-$Ga_2O_3$ film via atomic-scale STEM investigations.

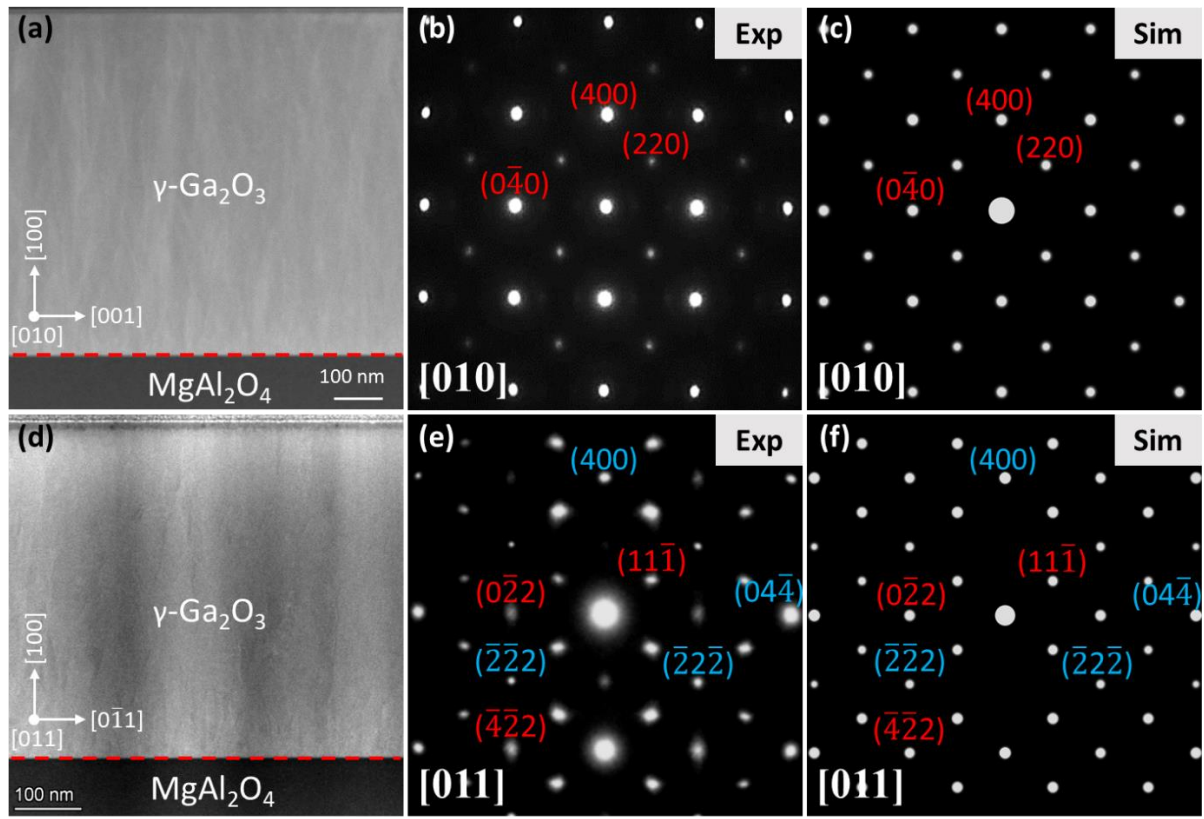

FIG. 2. Low-magnification cross-sectional bright-field STEM images of the film grown for two hours at 470 °C, acquired along (a) the [010] zone and (d) [011] zone. SAED patterns of the same γ-Ga$_2$O$_3$ layer acquired along (b) the [010] zone and (e) [011] zone. Simulated SAED patterns of γ-Ga$_2$O$_3$ acquired along (c) the [010] zone and (f) [011] zone.

Figure 3 presents the representative HAADF-EDX composition profile of the γ-Ga$_2$O$_3$ film grown at 470 °C. Five distinct areas (not shown) were selected and showed the same composition profile as shown in Fig. 3. Notably, a sharp contrast at the interface was observed and primarily attributed to an abrupt change in composition at the interface. The deviation in the Ga/O stoichiometry (~43%/~57%) can be attributed to the high uncertainty associated with the quantitative analysis for light elements like O because of low x-ray fluorescence yields and self-absorption. Negligible Mg and Al were detected within the γ-Ga$_2$O$_3$ film. Our previously published results on (i) high-temperature growth between 550 °C and 850 °C[54] and (ii) high-temperature post-growth annealing between 800 °C and 1000 °C[57] showed that Mg and Al interatomic diffusion began to occur at ~650 °C. This resulted in a gradual variation in the atomic concentration of all the elements at the interface and into the films. Thus, as noted above, the term "γ-Ga$_2$O$_3$" is used throughout this paper to specifically refer to the film grown at 470 °C to differentiate it from the terms "γ-phase" and "γ-Ga$_2$O$_3$ solid solutions" used in our previous publications.

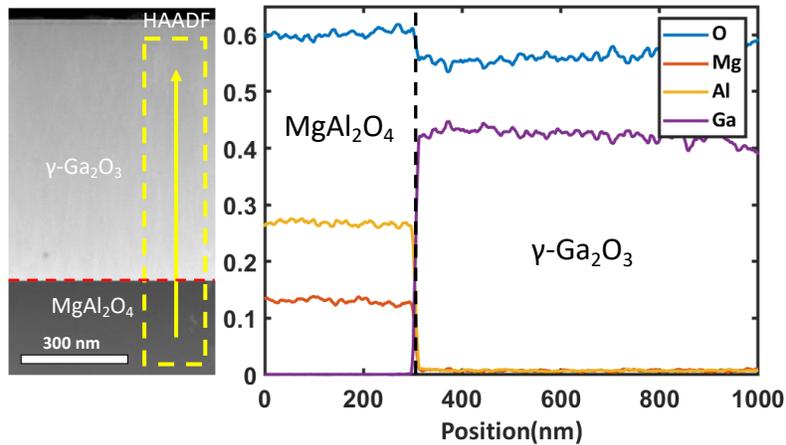

FIG. 3. HAADF-STEM image acquired along the [010] zone axis for the γ-Ga$_2$O$_3$ film grown at 470 ℃ (left) and corresponding integrated EDX composition profiles (right). The region and direction of the corresponding EDX profiles are highlighted in the yellow dashed box.

## II. Investigation on the planar defects inside the γ-Ga$_2$O$_3$ film

Figure 4 shows the HAADF-STEM results of the γ-Ga$_2$O$_3$ film grown for two hours at 470 ℃, acquired along the [011] zone. Three distinct features indicated by red arrows, yellow parallelograms, and blue parallelograms were observed throughout the entire film as shown in Figs. 4(a)-4(c). To gain a more detailed view, Figs. 4(d)-4(f) display the zoomed-in atomic-scale HAADF-STEM images of each feature. In HAADF-STEM, the contrast of the image is directly related to the column density and atomic number (Z) of the atoms and scales roughly as $Z^2$.[30,58,59] Consequently, columns of Ga atoms (Z = 31) are visible in the γ-Ga$_2$O$_3$ film, while the O atoms (Z = 8) are only visible if the column density for this element is sufficiently high. When viewed along the [011] zone axis, columns of Ga atoms do not overlap columns of O atoms, thus the contrast difference for the Ga sublattice observed in the atomic-resolution STEM images is purely due to the change in the column density of Ga atoms. Figure 4(d), from the regions marked by red arrows in Figs. 4(a)-4(c), shows a hexagonal atomic arrangement of ten columns of Ga atoms with an additional high-intensity single column of Ga atoms at the center of each hexagon. This arrangement aligns well with the ideal γ-Ga$_2$O$_3$ structure (the crystal model used here was the same as the one used in the previous section) viewed along the [011] zone, as depicted in Fig. 4(i). As can be seen in Fig. 4(i), each hexagon comprises six tetrahedrally occupied Ga atoms and five octahedrally occupied Ga atoms (four on the edge and one at the center). The column density of the Ga atoms at the center of each hexagon is twice compared with all the Ga atoms on the edge, resulting in a significantly higher contrast at the center in Fig. 4(d). Figures 4(e) and 4(f), corresponding to the regions marked by yellow parallelograms and blue parallelograms in Figs. 4(a)-4(c), respectively, reveal high-intensity diagonal stripes of Ga atoms and three atomic columns of Ga atoms with lower intensity between these high-intensity stripes. These types of Ga atomic arrangements are caused by a shift of the Ga sublattice in γ-Ga$_2$O$_3$ and the formation of planar defects.[35] Figures 4(j) and 4(k) show the overlaid structure of γ-Ga$_2$O$_3$ with proper lattice shift, aligning well with the observed features shown in Figs. 4(e) and 4(f). Further details regarding lattice shift will be discussed in conjunction with Fig. 5. An approximate 3 nm thick transition layer was observed between the film and the substrate shown in Fig. 4(c). Figure 4(g) shows a zoomed-in version of the pink area within the transition layer, presenting a regular hexagonal shape without any lattice shift. This type of feature matches well with the ideal γ-$Ga_2O_3$ structure shown in either Fig. 4(l) or 4(i). Figures 4(h) and 4(m) show the atomic-scale HAADF-STEM image of the $MgAl_2O_4$ substrate and its corresponding crystal structure viewed along the [011] zone, respectively. The transition layer was fully strained to match the underlying substrate, i.e., no lattice relaxation occurred. The lattice shift observed above the transition layer originated from the formation of antiphase boundaries due to the film relaxation.

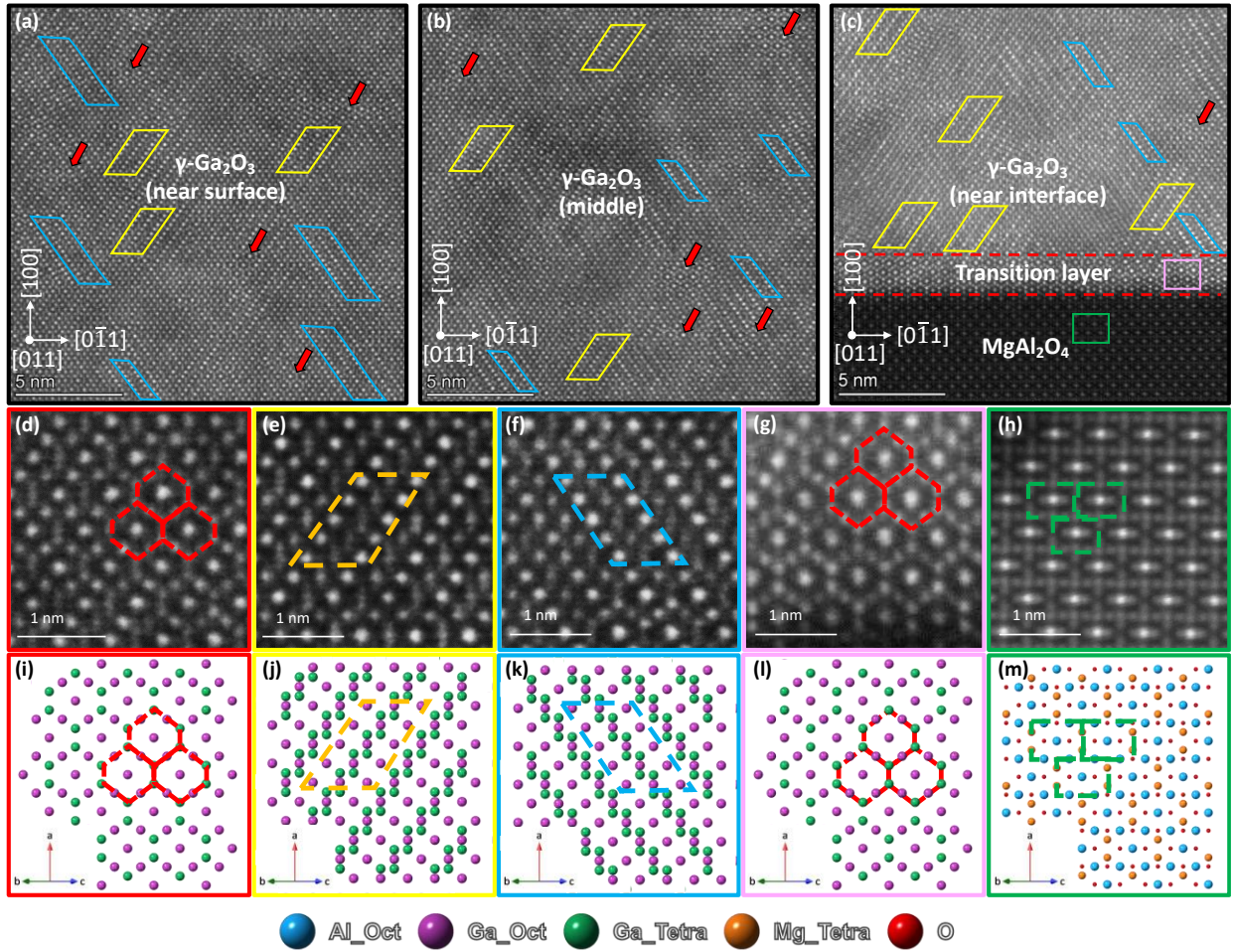

FIG. 4. High-magnification HAADF-STEM images along [011] zone for the γ-$Ga_2O_3$ film grown at 470 ℃ showing the representative area for (a) the top part of the film, (b) the middle part of the film and (c) the interface. Atomic-resolution HAADF-STEM images of the highlighted region of (d) red arrows, (e) yellow parallelograms, (f) blue parallelograms, (g) pink rectangle inside transition layer and, (h) green rectangle inside the substrate. (i)-(m) represent the corresponding schematic crystal model for (d)-(h), generated by CrystalMaker software.

Since its discovery in 1952, γ-$Ga_2O_3$ has been recognized as a cubic defective spinel structure.[24,40,60] However, due to its inherent disorder, the actual structure and the distribution of vacancies on the cation sites have been described in different ways in the literature. Ga atoms closely resemble Al atoms, which leads to similarities between $Ga_2O_3$ and $Al_2O_3$ polymorphs. α-, β-, κ-$Al_2O_3$ have been identified as prototypes of α-, β-, κ-Ga$_2$O$_3$. Similarly, γ-Ga$_2$O$_3$ is believed to have the same crystal structure as γ-Al$_2$O$_3$. Investigations of γ-Al$_2$O$_3$ have revealed the presence of planar defects, leading to broadening of specific diffraction peaks observed through XRD or SAED,[56,61] and similar to the results presented in Fig. 2(e). Here we apply the planar defects model, derived from lattice shifts observed within γ-Al$_2$O$_3$,[56,62,63] to describe the features observed in Figs. 4(e) and 4(f). The underlying assumption for the lattice shift model in generating planar defects is the preservation of atomic ordering within the cubic close-packed O sublattice while inducing a lattice shift within the cation sublattice (Al or Ga). Figure 5 illustrates the four types of planar defects produced by lattice shifts, which have been both experimentally and theoretically determined in γ-Al$_2$O$_3$. The glide plane is denoted by "(hkl)" and the corresponding shift vector by "[uvw]". Figures 5(a)-5(d) show the Ga sublattice for ideal γ-Ga$_2$O$_3$ before a lattice shift, as viewed along the [010] zone. The overlaid structures resulting from the indicated lattice shifts are shown in Figs. 5(a1)-5(d1) viewed along the [010] zone, and in Figs. 5(a2)-5(d2) viewed along the [011] zone. When viewing along the [011] zone, the overlaid structures in Figs. 5(a2) and (b2) match well with the actual HAADF-STEM image presented in Fig. 4(e), and Figs. 5(c2) and (d2) match those in Fig. 4(f). Further discussion of the HAADF-STEM images along the [010] zone and their comparison with Figs. 5(a1)-(d1) will be covered in reference to Fig. 6.

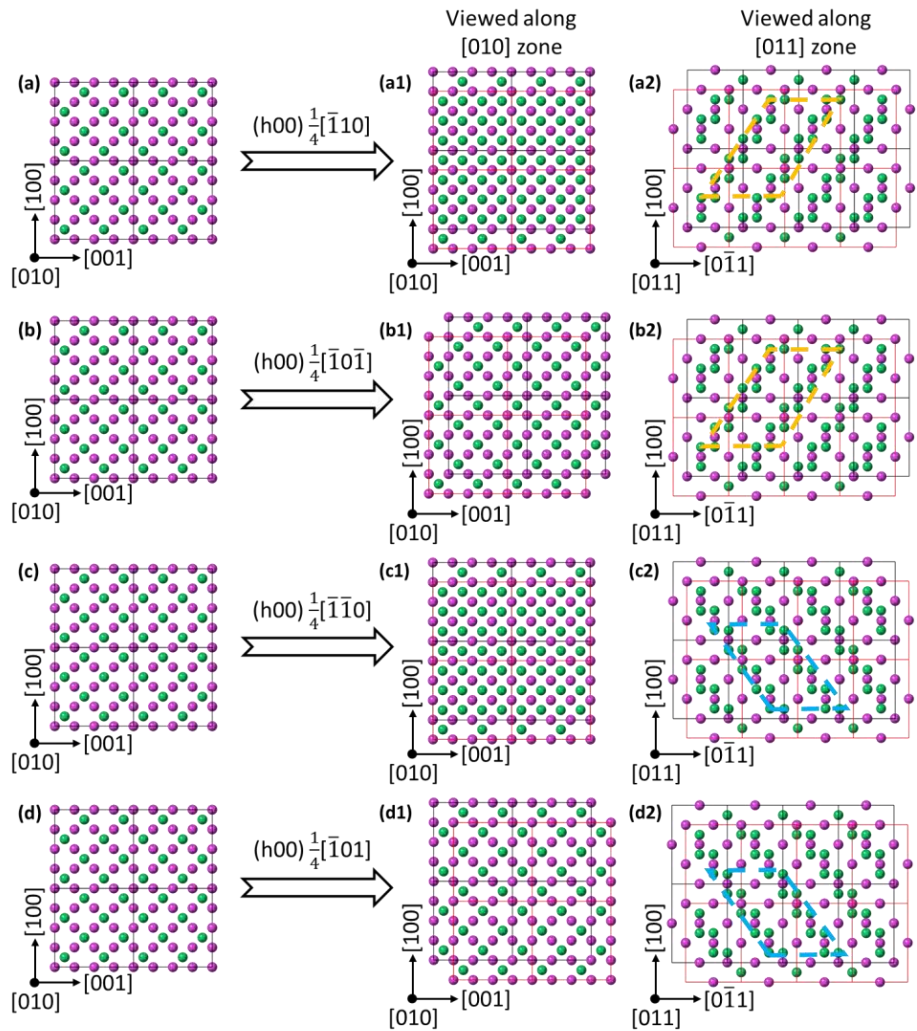

FIG. 5. Four types of planar defects (antiphase boundaries) experimentally and theoretically identified inside γ-Al$_2$O$_3$ and applied to γ-Ga$_2$O$_3$: (a) (h00) $\frac{1}{4}[\bar{1}10]$, (b) (h00) $\frac{1}{4}[\bar{1}0\bar{1}]$, (c) (h00) $\frac{1}{4}[\bar{1}\bar{1}0]$ and (d) (h00) $\frac{1}{4}[\bar{1}01]$. (a1)-(d1) show the overlaid crystal structure viewed along the [010] zone axis. (a2)-(d2) show the overlaid crystal structure viewed along the [011] zone axis.

Figure 6 shows the HAADF-STEM results of the γ-Ga$_2$O$_3$ film grown at 470 ℃, acquired along the [010] zone. Two distinct features indicated by red arrows and blue arrows were consistently observed throughout the film, as shown in Figs. 6(a)-6(c). To gain a more detailed view, Figs. 6(d) and 6(e) display the zoomed-in atomic-scale HAADF-STEM images of each feature. Although columns of Ga atoms partially overlap columns of O atoms when viewed along the [010] zone axis, the column density of the O atoms is the same before and after the lattice shift, indicating that the variation in the contrast for the Ga sublattice observed in the atomic-resolution STEM images is also purely due to the change in the column density of Ga atoms. Figure 6(d) acquired from the regions marked by blue arrows in Figs. 6(a)-6(c), shows a cubic atomic arrangement of Ga with high-intensity and low-intensity columns of Ga atoms occupying octahedral sites and tetrahedral sites, respectively. This arrangement aligns well with the overlaid γ-Ga$_2$O$_3$ structure interpreted by the lattice shift model presented in Figs. 5(a1) and 5(c1). Figure 6(e), corresponding to the regions marked by red arrows in Figs. 6(a)-6(c), reveals an ideal γ-Ga$_2$O$_3$ structure without any lattice shift or the overlaid structure produced by proper lattice shift, as presented in Figs. 5(b1) and 5(d1). Similar to the observation from the [011] zone, an approximate 3 nm thick transition layer was observed between the film and the substrate shown in Fig. 6(c). Figure 6(f) shows a magnified version of the region inside the transition layer, which reveals an ideal γ-Ga$_2$O$_3$ structure without lattice shift, as achieved by combining the knowledge acquired along the [011] zone at the interface. Figure 6(g) shows an atomic-scale HAADF-STEM image of the MgAl$_2$O$_4$ substrate. By comparing the in-plane lattice constant between the transition layer and the MgAl$_2$O$_4$ substrate, it was determined that the transition layer was fully strained to match the substrate.

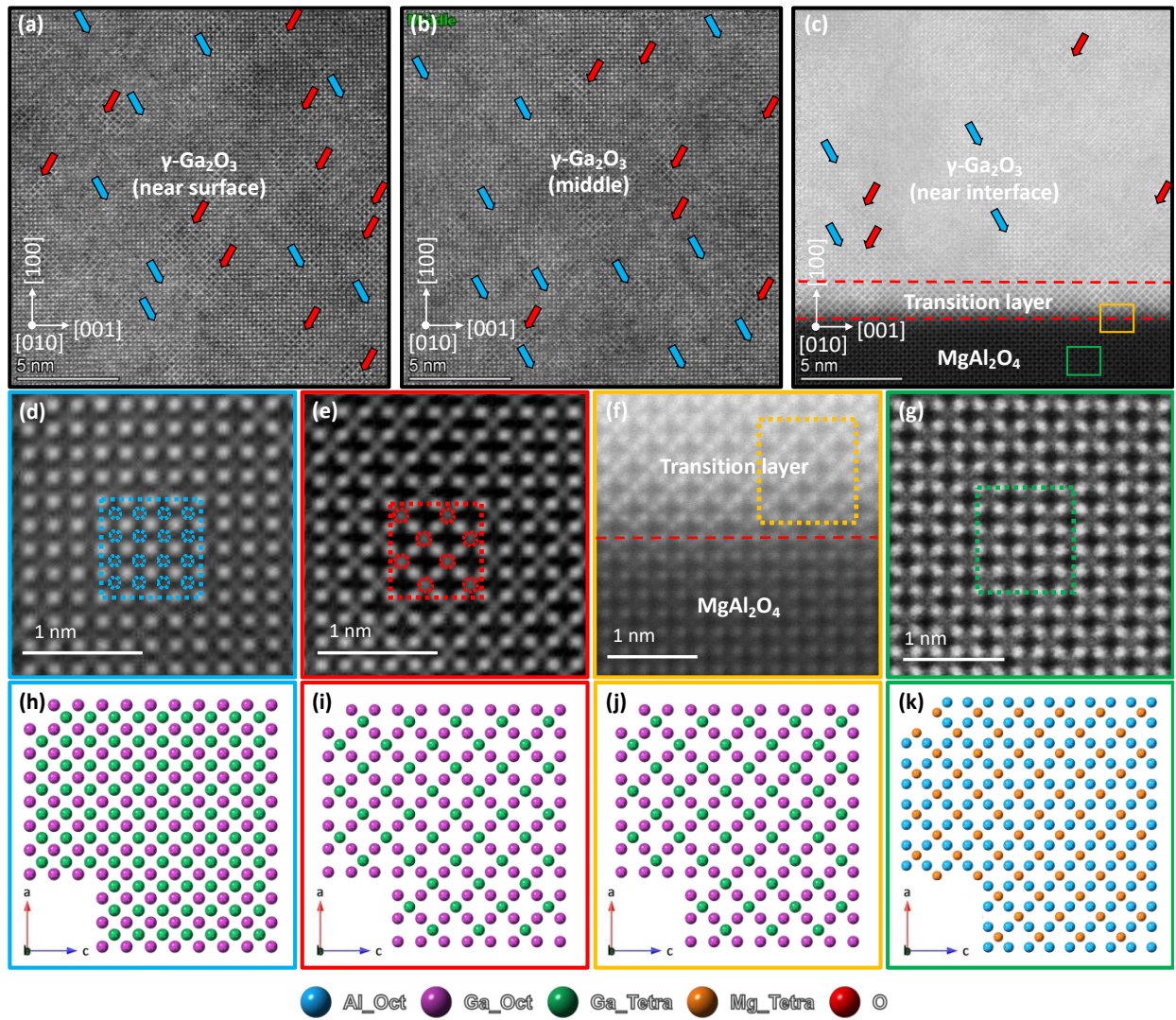

FIG. 6. High-magnification HAADF-STEM images along the [010] zone for the film grown at 470 °C showing representative areas for (a) the top part of the film, (b) the middle part of the film and (c) the interface. Atomic-resolution HAADF-STEM images of (d) region near a blue arrow, (e) region near a red arrow, (f) orange rectangle inside the transition layer, and (g) green rectangle inside the substrate. (h)-(k) represents the corresponding schematic crystal models for (d)-(g), generated using CrystalMaker software.

### III. Investigation of the relationship between γ-Ga₂O₃ and β-Ga₂O₃

The presence of "γ-phase" inclusions within β-$Ga_2O_3$ epitaxial layers and bulk materials has garnered significant attention, particularly in materials where various dopants such as Si, Ge, Sn, Mg, Al, or Sc have been introduced. The reported observations of the coexistence of two or more polymorphs require in-depth investigations of the possible epitaxial relationships among these polymorphs. For example, do the "γ-phase" inclusions within β-$Ga_2O_3$ exhibit a discernible preferential orientation with the surrounding matrix phase or are they randomly oriented? Two in-plane epitaxial relationships have been identified, namely, [010] β-Ga$_2$O$_3$ || [011] γ-Ga$_2$O$_3$ and [001] β-Ga$_2$O$_3$ || [011] γ-Ga$_2$O$_3$. It is noteworthy that even though the TEM samples were prepared along the [001] zone of β-Ga$_2$O$_3$, observations revealed the presence of both [001] and [010] β-Ga$_2$O$_3$ domains in the context of Al-alloyed β-Ga$_2$O$_3$.[32,34,38] The change in the domain orientation emanates from a transition in the growth plane from (010) to ($\bar{1}$02). The underlying mechanism for this 90° rotation (specifically from [001] β-Ga$_2$O$_3$ to [010] β-Ga$_2$O$_3$) has been experimentally and theoretically investigated.[32,34,38,64] Bhuiyan et al. concluded that this 90° rotation of β-Ga$_2$O$_3$ is due to the strain induced by a high Al concentration, which also promoted the growth of [011] γ-Ga$_2$O$_3$.[38] Chang et al. showed that the 90° rotation of Al alloyed β-Ga$_2$O$_3$ along the (310) β-Ga$_2$O$_3$ (approximately 36° with respect to the (010) growth plane), demonstrated that within these (310) planes, [001] β-Ga$_2$O$_3$ is either rotated to [010] β-Ga$_2$O$_3$ or converted to [011] γ-Ga$_2$O$_3$.[32] From the results of their DFT calculations, Wang et al. postulated that the 90° rotation is formed by the emergence of (010) β-Ga$_2$O$_3$ stacking faults at the interface which connect the [001] β-Ga$_2$O$_3$ and [010] β-Ga$_2$O$_3$ domains.[64] This connection subsequently gives rise to the (010)/($\bar{1}$02) interface. In the following paragraphs, the results and conclusions concerning (i) the epitaxial relationship between β-Ga$_2$O$_3$ domains and (ii) the epitaxial correlation that bridges β-Ga$_2$O$_3$ and γ-Ga$_2$O$_3$ determined in the present study are discussed.

Figure 7(a) is a HAADF-STEM micrograph showing a β-Ga$_2$O$_3$ inclusion (red box) within the γ-Ga$_2$O$_3$ film (blue box and beyond) grown at 470 °C. The image was acquired along the [011] zone of γ-Ga$_2$O$_3$. The areas within the boxes are magnified in Figs. 7(b) and 7(c). Figure 7(b) shows the superimposed structure of [011] γ-Ga$_2$O$_3$ on the projected STEM image, which confirms the structure of γ-Ga$_2$O$_3$ under this area. Figure 7(c) shows the superimposed structure of [010] β-Ga$_2$O$_3$ on the projected STEM image, confirming the structure of highly defective β-Ga$_2$O$_3$ in this area. The term "highly defective" is used to describe the observation of excessive interstitial Ga columns marked by the red dashed circles in Fig. 7(c). These additional interstitial Ga columns have been consistently noted in Si, Ge, Sn, and Al-doped β-Ga$_2$O$_3$.[33,34,36] The formation of these interstitial Ga columns has received experimental and theoretical validation,[33,34,36,65] substantiated using a model of interstitial-divacancy complexes. Additionally, the majority of these interstitial Ga columns marked, as indicated by the red dashed circles, exhibit a similar arrangement to the octahedrally positioned Ga atoms (purple atoms in Fig. 7(b)). This suggests that these additional Ga columns may have formed due to the overlap of the [010] β-Ga$_2$O$_3$ domain and the [011] γ-Ga$_2$O$_3$ domain. Given that β-Ga$_2$O$_3$ is embedded within the γ-Ga$_2$O$_3$ matrix, the possibility of an overlap between these two phases is quite significant. The presence of a β-Ga$_2$O$_3$ inclusion within the γ-Ga$_2$O$_3$ film is evident as revealed in Fig. 7(c). However, the extent of this inclusion and the limited presence of β-Ga$_2$O$_3$ in more than 30 regions within additional STEM images were insufficient for comprehensive STEM and XRD characterization. As such, an ex-situ post-annealing treatment at 600 °C in ambient air was conducted to promote the formation of a greater concentration of β-Ga$_2$O$_3$. A post-annealing treatment was used instead of growing films at higher temperatures. In the latter case, ($\bar{1}$12) oriented β-Ga$_2$O$_3$ also occurs in addition to the formation of (400) oriented β-Ga$_2$O$_3$. In β-Ga$_2$O$_3$ the ($\bar{1}$12) plane has been determined to possess a higher surface energy than the (400) plane.[66] Due to the intrinsic similarities between the ($\bar{1}$12) and the (400) planes of β-Ga$_2$O$_3$, the co-existence of both orientations is favorable during growth, whereas this is less likely during post-annealing.

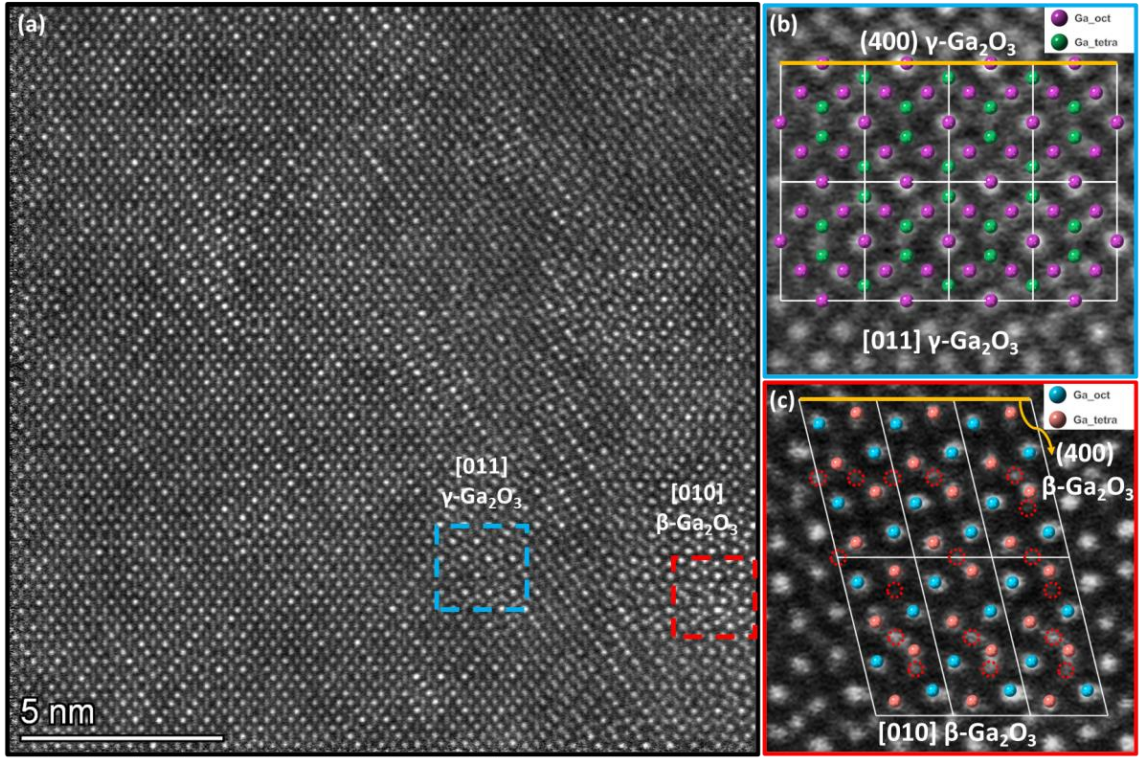

FIG. 7. High-magnification HAADF-STEM images along [010] zone for the film grown at 470 °C showing (a) a β-$Ga_2O_3$ inclusion inside the γ-$Ga_2O_3$, (b) magnified region of γ-$Ga_2O_3$ with the overlaid schematic crystal model of γ-$Ga_2O_3$ and (c) magnified region of β-$Ga_2O_3$ with the overlaid schematic crystal model of β-$Ga_2O_3$.

Figure 8 shows the rotational domains of the β-$Ga_2O_3$ within the γ-$Ga_2O_3$ film as well as the corresponding epitaxial relationship between β-$Ga_2O_3$ and γ-$Ga_2O_3$ as derived from XRD ϕ-scans. Figure 8(a) reveals recrystallization of β-$Ga_2O_3$ within the γ-$Ga_2O_3$ film annealed at 600 °C in ambient air for 1 hour. The peak for (400) β-$Ga_2O_3$ is barely detectable before annealing; a broader (400) β-$Ga_2O_3$ peak is visible after heat treatment. The (400) γ-$Ga_2O_3$ peak has slightly shifted to a higher 2θ value, indicative of ongoing film relaxation. Figure 8(b) shows the XRD ϕ-scan of {$40\bar{1}$} β-$Ga_2O_3$ and {440} γ-$Ga_2O_3$ for the post-annealed film. Four peaks with a 90° separation for each indicates the presence of four in-plane rotational domains in β-$Ga_2O_3$. The [010] β-$Ga_2O_3$ inclusion shown in Figure 7 contains one of these domains. The epitaxial relationship between β-$Ga_2O_3$ and γ-$Ga_2O_3$ is presented in Fig. 8(c). The presence of the in-plane rotational domains of β-$Ga_2O_3$ can be attributed to the four-fold symmetry of the (400) plane of γ-$Ga_2O_3$. Moreover, the domain boundaries between the 180° rotated domains of either 1 and 3 or 2 and 4 stem from twin boundaries formed on the (100) plane of β-$Ga_2O_3$. These twin domains, achieved via a boundary plane mirror operation followed by a 0.5c displacement in the [001] direction, are a common occurrence in β-$Ga_2O_3$ homoepitaxial films. DFT calculations validate the lower formation energy of twin boundaries on the (100) plane of β-$Ga_2O_3$ relative to other grain boundaries, thereby providing insight into the origin of these 180° in-plane rotational domains.[64]

The initiation of the 90° rotation from [001] β-$Ga_2O_3$ to [010] β-$Ga_2O_3$, observed alongside [011] γ-$Ga_2O_3$ inclusions when Al is introduced was also considered. Using the evidence presented above, for the epitaxial relationship between β-$Ga_2O_3$ and γ-$Ga_2O_3$ (i.e., <010> β-$Ga_2O_3$ || <011> γ-$Ga_2O_3$), we postulate

that the formation of the 90° rotation of β-Ga$_2$O$_3$ may be enhanced by the presence of γ-Ga$_2$O$_3$. As γ-Ga$_2$O$_3$ possesses a higher symmetry than β-Ga$_2$O$_3$, the presence of the rotational domains becomes virtually inevitable. As illustrated in Fig. 8(c), visualizing γ-Ga$_2$O$_3$ along the [011] zone axis reveals that two domains of β-Ga$_2$O$_3$ are well aligned with respect to the [011] γ-Ga$_2$O$_3$ and constitute ± [001] domains (domains 1 and 3) and ± [010] domains (domains 2 and 4) of β-Ga$_2$O$_3$. Considering the role of Al in stabilizing the formation of the γ-phase,[67] we further postulate that the Al-induced emergence of the γ-phase ultimately triggers the 90° rotation in the β-phase.

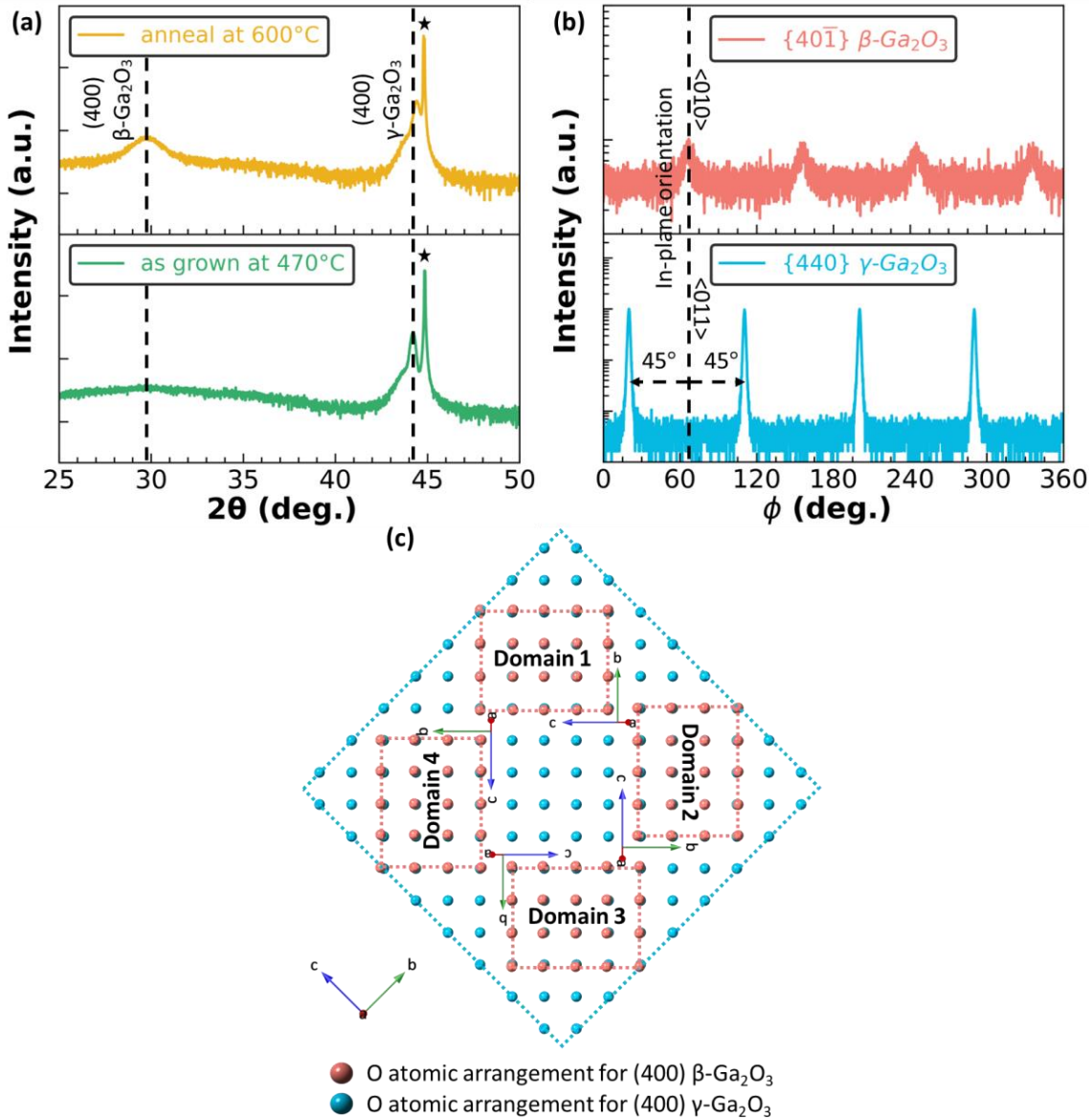

FIG. 8. (a) XRD 2θ-ω patterns (log scale) for (i) a γ-Ga$_2$O$_3$ film grown on MgAl$_2$O$_4$ at 470 °C and (ii) the same film post-annealed ex-situ at 600 °C in ambient air for 1 hour. Black stars indicate the (400) plane of MgAl$_2$O$_4$. (b) XRD φ-scan of {40$\bar{1}$} β-Ga$_2$O$_3$ and {440} γ-Ga$_2$O$_3$, respectively, for the ex-situ post-annealed sample shown in Fig. 8(a). (c) Schematic of the epitaxial relationship between β-Ga$_2$O$_3$ and γ-Ga$_2$O$_3$ derived from XRD φ-scans.

## SUMMARY

This research revealed (i) a narrow window for the successful growth via MOCVD of thin films of γ-$Ga_2O_3$ on (100) $MgAl_2O_4$ substrates, (ii) the formation of a significant density of antiphase boundaries within γ-$Ga_2O_3$, and (iii) the nature of the epitaxial relationship between γ-$Ga_2O_3$ and β-$Ga_2O_3$. Specifically, nominally phase pure γ-$Ga_2O_3$ was grown at 470 ℃; whereas, either a mixture of γ-$Ga_2O_3$ and β-$Ga_2O_3$ or only β-$Ga_2O_3$ formed at higher temperatures ranging from 500 ℃ to 530 ℃. Integration of the diffusion profiles for Mg and Al acquired using EDX, showed the absence of interatomic diffusion of these elements into the film grown at 470 ℃. Atomic resolution STEM results showed that the antiphase boundaries produced lattice shifts in the Ga sublattice within the entire γ-$Ga_2O_3$ film grown at 470 ℃. Twin boundaries (180° rotational domains) as well as 90° rotational domains were observed in the β-$Ga_2O_3$ inclusions that formed within the γ-$Ga_2O_3$ matrix. The results of a comprehensive investigation of the epitaxial relationship between these two phases were presented and reconciled with existing reports concerning γ-phase inclusions containing various dopants. It was hypothesized that the introduction of dopants such as Si, Ge, Sn, Mg, Al, and Sc into β-$Ga_2O_3$, locally stabilizes the "γ-phase" based on their preferred site occupancy. This effectively favors the formation of the "γ-phase" or γ-$Ga_2O_3$ solid solutions over β-$Ga_2O_3$. In contrast, in the absence of such dopants, pure γ-$Ga_2O_3$ remains the least stable phase, as indicated by its narrower growth window and lower growth temperatures, when using MOCVD growth techniques, than other $Ga_2O_3$ polymorphs. Further in-situ TEM/STEM investigations on the phase transition from γ-$Ga_2O_3$ to β-$Ga_2O_3$ are ongoing and will be fully discussed in a separate paper.

## Acknowledgments

This material is based upon work supported by the Air Force Office of Scientific Research (Program Manager, Dr. Ali Sayir) under Award No. FA9550-21-1-0360, and by II-VI Foundation. The use of the Materials Characterization Facility at Carnegie Mellon University was supported by Grant No. MCF-677785. J. T. and K. J. would also like to acknowledge the Neil and Jo Bushnell Fellowship in Engineering for additional support.

## Conflict of Interest

The authors have no conflict to disclose.

## Data Availability

The data that support the findings of this study are available from the corresponding author upon reasonable request.